%% file: ichep04-mu.tex
\documentclass{ws-procs10x7}
\input{babarsym}
\input{definitions}

\begin{document}

\title{$CP$ Violation and the CKM Matrix: \\ Impact of the Asymmetric B Factories}

\author{Muriel Pivk}

\address{{\small (on behalf of the CKMfitter Group)} \\ CERN, PH-EP Department, 1211 Geneva 23, Switzerland \\E-mail: muriel.pivk@cern.ch}

\twocolumn[\maketitle\abstract{An up-to-date profile of the CKM matrix is presented, with emphasis on the interpretation of recent $CP$-violation results from the $B$ factories. The apex of the Unitarity Triangle is determined by a global CKM fit. A study is performed to probe the dynamics of $B$ decays into $\pi\pi$, $K\pi$, $\rho\pi$ and $\rho\rho$ within two theoretical frameworks. A model-independent investigation of New Physics effects in $\Bz\Bzb$ mixing is given.}]
 
\section{Introduction}

\vspace{-0.2cm}
The evolution of the knowledge of the Cabibbo-Kobayashi-Maskawa (CKM) matrix\cite{bib:C,bib:KM} profile has been very important in the last few years, most notably thanks to the performances of the $B$ factories KEK-B and \pep2\ and their two detectors Belle and BABAR respectively. The development of the numerical analyser package CKMfitter\cite{bib:pap1} is dedicated to the comprehensive study of the CKM-matrix constraints\cite{bib:ThePap}.


The CKMfitter package features (among others) the {\em frequentist} approach \rfit. The theoretical errors are interpreted as allowed ranges and no other {\em a priori} information is assumed when there is none available. The global CKM analysis pursues three different goals. The first one is to probe the SM by quantifying the agreement between the data and the theory. The second goal is, within the SM, to perform a careful metrology of the parameters. Finally, a search for specific signs of New Physics (NP) is performed within extended theoretical frameworks.

\section{The global CKM fit}

\begin{figure}
\epsfxsize170pt
\figurebox{120pt}{160pt}{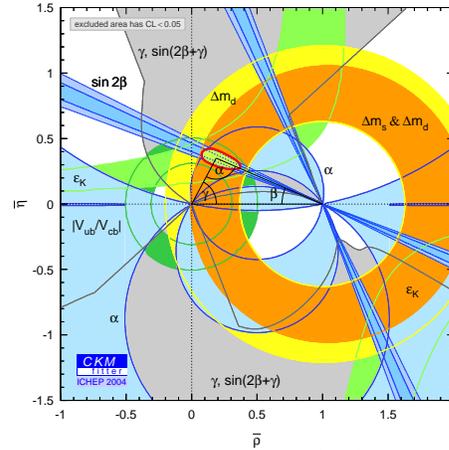}

\vspace{-0.3cm}
\caption{\em Confidence levels in the $\rhoeta$ plane for the present standard CKM fit. The shaded areas indicate the regions of $\ge 5\%$ CLs. For $\stwob$ the $\ge 32\%$ and $\ge 5\%$ CL constraints are shown.}
\label{fig:glo}
\end{figure}

\vspace{-0.2cm}
The present status of the constraints on the Unitarity Triangle (UT) is represented Fig.~{\ref{fig:glo}}. This corresponds to the overall constrained CKM fit, denoted {\em standard CKM fit} in the following. The inputs\cite{bib:ThePap}, given in Table~\ref{tab:inputs}, can all be considered quantitatively under control.

\begin{table}[h!]
\begin{tabular}{|c|c|} 
\hline 
Observables & Value $\pm$ error(s) \\
\hline \hline
$\Vus$ & $0.2228 \pm0.0039 \pm 0.0018$ \\
$\Vud$ & $0.9735^{+0.0005}_{-0.0001} {}^{+0.0004}_{-0.0003}$ \\
$\Vub$ & $(3.90 \pm 0.08 \pm 0.68) \times 10^{-3}$ \\
$\Vcb$ incl. & $(42.0 \pm 0.6 \pm 0.8) \times 10^{-3}$ \\
$\Vcb$ excl. & $(40.2^{+2.1}_{-1.8}) \times 10^{-3}$ \\
\hline
$\mid \epsk \mid$ & $(2.282 \pm 0.017) \times 10^{-3}$ \\
\hline
$\dmd$ & $(0.502 \pm 0.006) \ps^{-1}$ \\
$\dms$ & Amplitude spectrum \\
\hline
$\stwob$ & $0.726 \pm 0.037$ \\
$\alpha$ & $(103^{+9}_{-10})^\circ$ \\
\hline
\end{tabular}
\caption{\em Inputs to the standard CKM fit.}
\label{tab:inputs}
\end{table}
\begin{table}
\begin{center}
\begin{tabular}{|c|c|} 
\hline 
Quantity & Value $\pm$ error(s) \\
\hline \hline
$\rhobar$ & $0.22^{+0.06}_{-0.13}$ \\
$\etabar$ & $0.334^{+0.030}_{-0.029}$ \\
\hline
$\stwob$ & $0.67^{+0.18}_{-0.08}$ \\
$\alpha$ & $(98 \pm 16)^\circ$ \\
$\gamma$ & $(57 \pm 9)^\circ$ \\
\hline
\end{tabular}
\caption{\em Results from the standard CKM fit for $\rhobar$ and $\etabar$. The three other results are obtained the variable being excluded from the fit.}\label{tab:res}
\end{center}
\end{table}
The fit is dominated by the precision measurement of $\stwob$ from the $B$ factories. The recent measurements of $\alpha$ from the time-dependent $CP$-violating asymmetries in $B \to \pi\pi$, $\rho\rho$ and $\rho\pi$ decays (Section~\ref{sec:alpha}) are included for the first time. Two additional constraints are shown Fig.~\ref{fig:glo}, one for $\gamma$ from $\Bp \to$ $\Dz (\KS \pip \pim)$ $\Kp$\cite{bib:bellegamma} and one for $\sin(2\beta+\gamma)$ from $\Bz \to$ $D^{\star \pm} \pi^{\mp}$\cite{bib:hfag}, but they are, so far, insufficient to improve the knowledge of the apex of the UT. However, they play, with $\alpha$, a more significant role in constraining NP~(Section~\ref{sec:NP}). The results obtained from the standard CKM fit are given Table~\ref{tab:res}.

\begin{figure}
\epsfxsize170pt
\figurebox{120pt}{160pt}{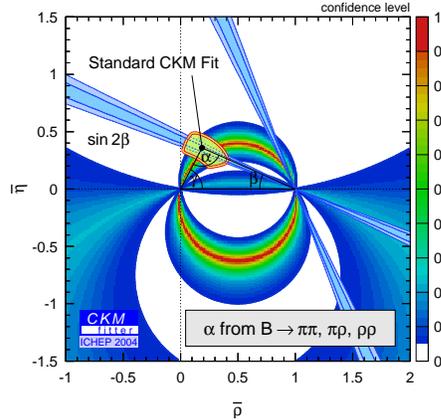}

\vspace{-0.3cm}
\caption{\em Confidence levels in the $\rhoeta$ plane for the constraints from $\stwob$ and $\alpha$. The standard CKM fit, excluding both of them is superimposed in order to show the impact of the $B$ factories.}
\label{fig:glo-s2b-a}
\end{figure}

Figure~\ref{fig:glo-s2b-a} illustrates the constraints from $\stwob$ and $\alpha$ on the $\rhoeta$ plane as measured by the $B$ factories. Superimposed is the standard CKM fit without these two constraints.

\section{Charmless $B$ decays}

\vspace{-0.2cm}
Different weak phases must be considered in the charmless $B$ decay analyses, in contrast with most of the charmed ones, where the amplitudes contain only one weak phase. This makes the extraction of the experimental observables more difficult and the use of theoretical assumptions becomes necessary. In the following, two theoretical frameworks, the model-independent isospin analysis\cite{bib:iso} and QCD Factorisation\cite{bib:qcdfa} (QCD FA) are considered.

\subsection{Constraints on $\alpha$}\label{sec:alpha}

\begin{figure}
\epsfxsize170pt
\figurebox{120pt}{160pt}{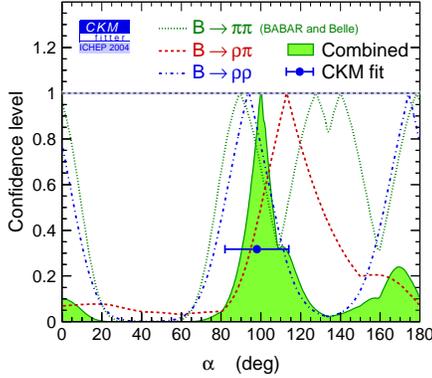}

\vspace{-0.8cm}
\caption{\em Confidence level for~$\alpha$ obtained from the three charmless $B$ decays $B \to \pi\pi$, $\rho\rho$ and $\rho\pi$. The combined fit (shaded) and the standard CKM fit (dot with error bar) are also shown.}
\label{fig:CLalpha}
\end{figure}

\vspace{-0.1cm}
Three different decays lead at present to the measurement of $\alpha$. The $\B \to \pi\pi$ decays, using a triangular isospin relation involving the first measurement of the $CP$-violating asymmetry $\Coo$\cite{bib:crin} from $\Bztopizpiz$, start to help identifying the mirror solutions of~$\alpha$. The large contribution of penguin diagrams does not allow yet a precise constraint of this angle. The $\B \to \rho\rho$ decays\cite{bib:carlo}, similar to the $\B \to \pi\pi$ ones, but containing much smaller penguins, give a more precise measurement of~$\alpha$. The $\B \to \rho\pi$ decays\cite{bib:rhopi}, using a Dalitz analysis (the isospin relation is pentagonal and therefore imprecise) provide a two-fold ambiguity concerning the value of~$\alpha$. The results of these three channels being consistent, they are combined to yield:
\begin{equation}
\alpha = (103^{+9}_{-10})^\circ ~.
\end{equation}
The confidence level on~$\alpha$ for the three decays separately and combined are shown Fig.~\ref{fig:CLalpha}.

\subsection{Standard Model test}

\begin{figure}
\epsfxsize170pt
\figurebox{120pt}{160pt}{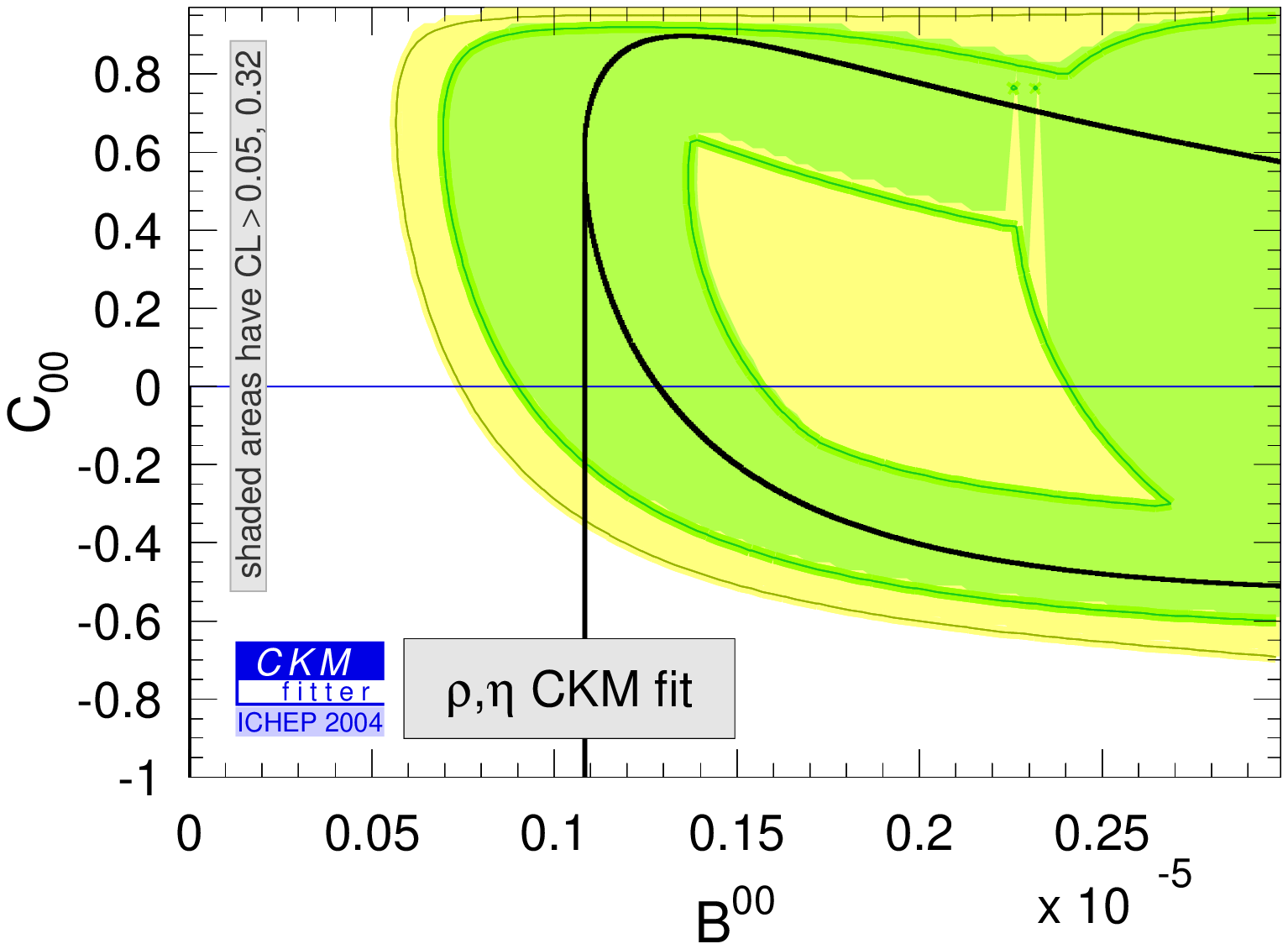}

\vspace{-0.3cm}
\caption{\em Confidence level in the ($\Broo,\Coo$) plane at an integrated luminosity of $1~\invab$. The curve superimposed represents the function $\Coo(\Broo)$.}
\label{fig:cooboo}
\end{figure}

\vspace{-0.1cm}
Charmless $B$ decays also provide new ways of testing the SM. Using the $\Bz \to \piz\piz$ decay and, in order to avoid to deal with the mirror solutions for the angle~$\alpha$ (Fig.~\ref{fig:CLalpha}), it is appropriate to consider the $(\Broo,\Coo)$ plane\cite{bib:boonote}. It is represented in Fig.~\ref{fig:cooboo} considering the present world average central values, but for errors extrapolated to a luminosity of $1~\invab$. The figure stresses the fact that there is still a long way to go to exclude the SM, since only a small fraction of the plane can be excluded.

\subsection{Adding QCD factorisation}\label{sec:qcdfa}

\begin{figure}
\begin{center}
\mbox{\epsfig{file=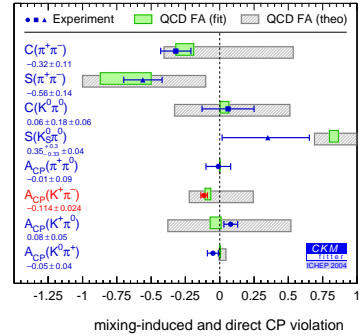,width=0.7\linewidth}}
\mbox{\epsfig{file=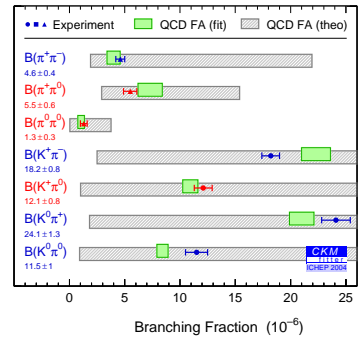,width=0.7\linewidth}}

\vspace{-0.5cm}
\caption{\em Comparison of the results from the global QCD FA fit to $\B \to \pi\pi$, $K\pi$ data to the unconstrained QCD FA predictions and to experimental results. The $CP$-violating asymmetries are displayed on the left and the branching ratios on the right.}
\label{fig:qcdfa}
\end{center}
\end{figure}

\vspace{-0.1cm}
Combining the experimental results on $\B \to \pi\pi$ and $\B \to K\pi$ decays with the QCD FA theoretical framework within a global fit, predictions can be made for each observable, ignoring the measurement available for it in the fit. The results are then unbiased by this treatment. The standard CKM fit is included as well. The predictions for the branching ratios and the $CP$ asymmetries are displayed in Fig.~\ref{fig:qcdfa} together with the experimental results. The predictions of the full QCD FA fit are in agreement with the measurements, modulo final state radiative corrections\cite{bib:TheseMu}, which, when accounted for in the experimental analyses, may affect significantly some of the channels, and with the exception of the branching fractions for $\Kp\pim$ and $\Kz\piz$~\cite{bib:kpi}.

\section{New Physics in $\BzBzb$ mixing}\label{sec:NP}

\vspace{-0.2cm}
So far, the SM is able to accomodate the data within the present experimental uncertainties so that it does not seem relevant to introduce contributions from physics beyond the SM. However, NP contributions are not necessarily absent and it is worth investigating by how much specific NP parameters are constrained. The NP analysis performed here proceeds as follows: the observables expected to be dominated by the SM contributions are used to construct a model-dependent UT followed by a constrained fit on NP contributions in $\BzBzb$ mixing. 

\begin{figure}
\epsfxsize160pt
\figurebox{120pt}{160pt}{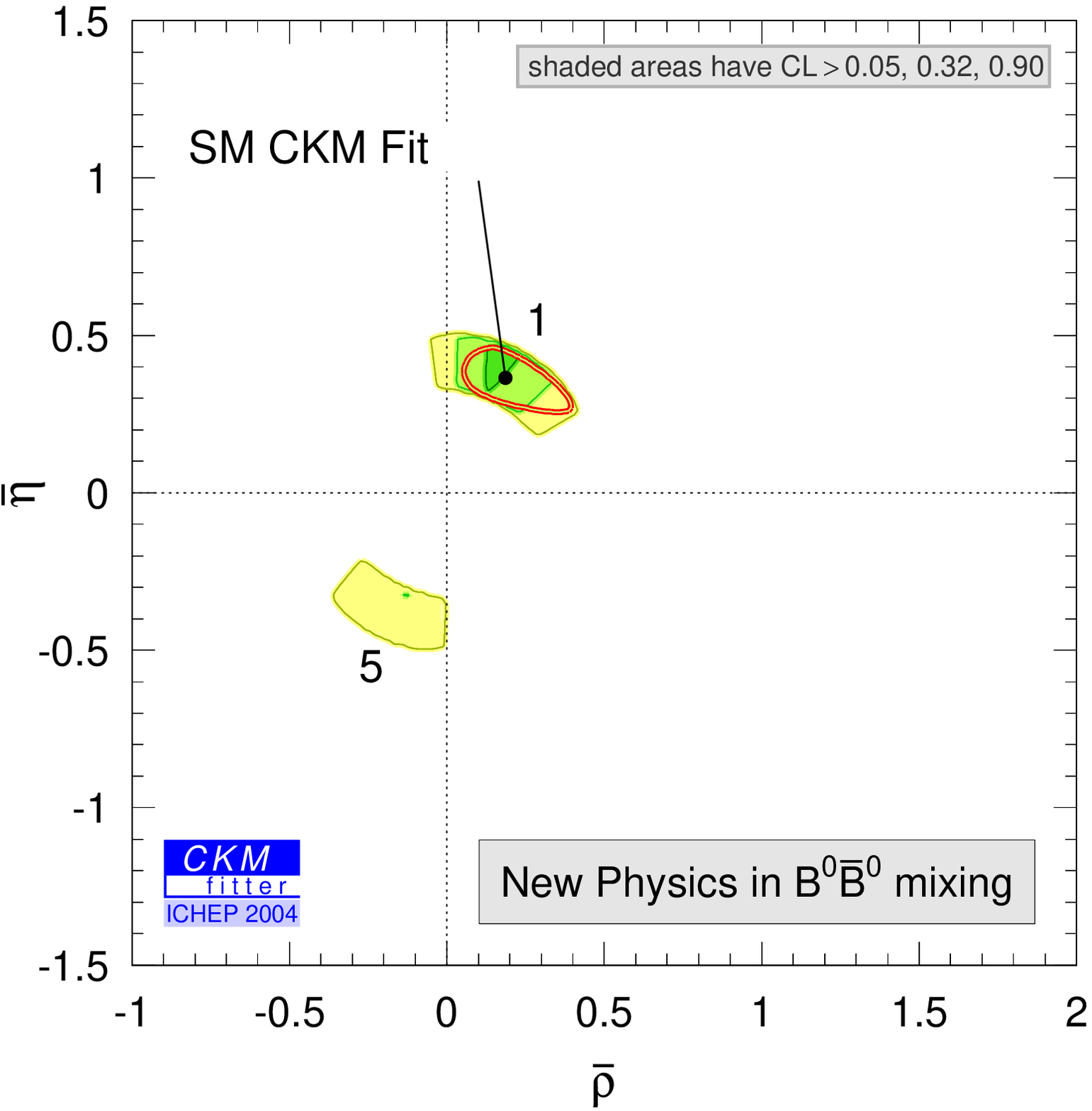}

\vspace{-0.3cm}
\caption{\em Constraints in the $\rhoeta$ plane from the fit in the framework of NP in $\BzBzb$ mixing.}
\label{fig:NPSM}
\end{figure}

\begin{figure}
\epsfxsize160pt
\figurebox{120pt}{160pt}{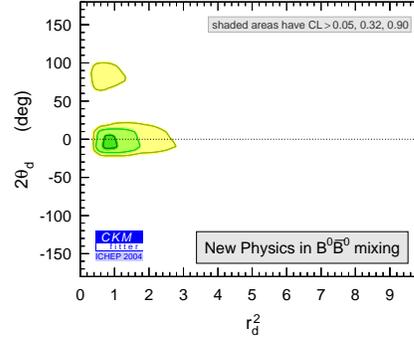}

\vspace{-0.3cm}
\caption{\em Confidence level in the $(r_{d}^2,2\theta_{d})$ plane obtained as a result of the global CKM fit that includes NP in $\BzBzb$ mixing.}
\label{fig:NP}

\end{figure}

To have a parameterisation which is as model independent as possible, two parameters, $r_{d}^2$ and $2\theta_{d}$, defined by\cite{bib:NP}:
\begin{equation}
  r_{d}^2\,e^{i 2\theta_{d}} \:=\:
  \frac{\langle \Bz | {\mathcal{H}}_{\rm eff}^{\rm full}| \Bzb\rangle}
  {\langle \Bz | {\mathcal{H}}_{\rm eff}^{\rm SM}| \Bzb\rangle}~,
\end{equation}
are introduced. Both CKM $\rhoeta$ and NP $(r_{d}^2,2\theta_{d})$ planes can be constrained, as shown Figs.~\ref{fig:NPSM} and~\ref{fig:NP} respectively.
In the $\rhoeta$ plane, the solution with the largest confidence level is consistent with the SM. In the $(r_{d}^2,2\theta_{d})$ plane, the SM values for these parameters are favored ($r_{d}^2 = 1$ and $2\theta_{d} = 0$).

%

\section*{Acknowledgments}

\vspace{-0.3cm}
I am happy to thank all the people who made this conference a very nice adventure !


\end{document}

%% file: definitions.tex
\def\slashchar#1{\setbox0=\hbox{$#1$}   
  \dimen0=\wd0                          
  \setbox1=\hbox{/} \dimen1=\wd1
  \ifdim\dimen0>\dimen1         
  \rlap{\hbox to \dimen0{\hfil/\hfil}}  
  #1                                    
  \else                                 
  \rlap{\hbox to \dimen1{\hfil$#1$\hfil}}
  /                                      
  \fi}                                         %

\def\Broo{\BR^{00}}
\def\Coo{C_{00}}

\newcommand{\beq}{\begin{equation}}
\newcommand{\eeq}{\end{equation}}
\newcommand{\beqn}{\begin{eqnarray}}
\newcommand{\eeqn}{\end{eqnarray}}

\def\dmd{\Delta m_d}
\def\2{\black 2}

\def\epsk{\epsilon_K}
\def\dms{\Delta m_s}

\def\su3bris{SU(\slashchar{3})}

\newcommand\rfit{{\em R}fit}

\def\rhoeta{(\bar\rho,\bar\eta)}

\def\L02{{\{L_0,L_2\}}}

\def\sinflb2{\sin^2\theta_B^\ast}

\def\dz{\Delta z}

\def\NP{{\em Nucl. Phys.}}

\def\ea{{\em et al.}}